\documentclass{aa}  

\usepackage{graphicx}
\usepackage{txfonts}
\newcommand{\masyr}{{\,\mathrm{mas}\,\mathrm{yr}^{-1}}}
\newcommand{\kms}{{\,\mathrm{km}\,\mathrm{s}^{-1}}}

\begin{document} 


   \title{$Gaia$ DR2 reveals a very massive runaway star ejected from R136}

\titlerunning{Massive runaways in 30 Dor}
\authorrunning{D.J. Lennon et al.}

   \author{D.J. Lennon\inst{1}
   \and
          C.J. Evans \inst{2}
          \and
          R.P. van der Marel\inst{3,4}
          \and
          J. Anderson\inst{3}
          \and
          I. Platais\inst{4}
          \and
          A. Herrero\inst{5,6}
          \and
          S.E. de Mink\inst{7}
          \and
          H. Sana\inst{8}
          \and
          E. Sabbi\inst{3}
          \and
          L.R. Bedin\inst{9}
          \and
          P.A. Crowther\inst{10}
          \and
          N. Langer\inst{11}
          \and
          M. Ramos Lerate\inst{12}
          \and
          A. del Pino\inst{3}
          \and
          M. Renzo\inst{7}
          \and
          S. Sim{\'o}n-D{\'i}az\inst{5,6}
          \and
          F.R.N. Schneider\inst{13}
          }
{
   \institute{ESA, European Space Astronomy Centre, Apdo. de Correos 78,
E-28691 Villanueva de la Ca\~nada, Madrid, Spain
             \email{danny.lennon@sciops.esa.int}
        \and
        UK Astronomy Technology Centre, Royal Observatory Edinburgh,
Blackford Hill, Edinburgh, EH9 3HJ, United Kingdom      
 \and
           Space Telescope Science Institute, 3700 San Martin Drive, Baltimore, MD 21218, USA
           \and
           Department of Physics and Astronomy, Johns Hopkins University, 3400 North Charles Street, Baltimore, MD 21218, USA
                      \and
           Instituto de Astrof\'{i}sica de Canarias, 38200 La Laguna, Tenerife, Spain
           \and
           Departamento de Astrof{\'i}sica, Universidad de La Laguna, E-38205, La Laguna, Tenerife, Spain
           \and
           Anton Pannekoek Institute for Astronomy, University of Amsterdam, Science Park 904, 1098 XH, Amsterdam, The Netherlands
           \and
           Institute of astrophysics, KU Leuven, Celestijnlaan 200D, 3001, Leuven, Belgium
           \and
INAF-Osservatorio Astronomico di Padova, Vicolo dell'Osservatorio 5, I-35122 Padova, Italy
\and
Department of Physics and Astronomy, Hicks Building, Hounsfield Road, University of Sheffield, Sheffield S3 7RH, United Kingdom
\and
Argelander-Instit\"ut f\"ur Astronomie, Universit\"at Bonn, Auf dem H\"ugel 71, 53121, Bonn, Germany           
         \and
            VitrocisetBelgium for ESA, European Space Astronomy Centre, Apdo. de Correos 78,
E-28691 Villanueva de la Ca\~nada, Madrid, Spain
\and
Department of Physics, University of Oxford, Denys Wilkinson Building, Keble Road, 
Oxford OX1 3RH, United Kingdom
              }

   \date{Received ; accepted}

 
  \abstract{A previous spectroscopic study identified the very massive 
  O2\,III star VFTS~16 in the Tarantula Nebula as a runaway star based on its peculiar line-of-sight velocity. We use the  {\em Gaia} DR2 catalog to measure the relative
  proper motion of VFTS~16 and nearby bright stars to test if this star might
  have been ejected from the central cluster, R136, via dynamical ejection.
   We find that the position angle and magnitude of the relative proper motion
   ($0.338\pm0.046\masyr$,  or approximately $80\pm11$ $\kms$)
   of VFTS~16 are consistent with ejection from R136 approximately $1.5\pm 0.2$ Myr
   ago, very soon after the cluster was formed. There is some tension with the presumed
   age of VFTS~16 that, from published stellar parameters, cannot be greater than
   $0.9^{+0.3}_{-0.2}$ Myr.
   Older ages for this star would appear to be
   prohibited due to the absence of He\,{\sc i} lines in its optical spectrum,
   since this sets a firm lower limit on its effective temperature. 
   The dynamical constraints may imply an unusual evolutionary history for this object,
   perhaps indicating it is a merger product.
   $Gaia$ DR2 also confirms that another very massive star in the Tarantula Nebula, 
   VFTS~72 (alias BI~253; O2\,III-V(n)((f*)),  is also a runaway on the basis of its proper
   motion as measured by $Gaia$. While its tangential proper motion 
   ($0.392\pm0.062\masyr$ or $93\pm15$ $\kms$) 
   would be consistent with dynamical ejection from R136 approximately 1 Myr ago, 
   its position angle 
   is discrepant with this direction at the $2\sigma$ level.
   From their $Gaia$ DR2 proper motions we conclude that 
    the two $\sim$100M$_{\odot}$ O2 stars,
   VFTS~16 and VFTS~72, are fast runaway stars, 
   with space velocities of around 100 $\kms$ relative to R136
   and the local massive star population. 
   The dynamics of VFTS~16 are consistent with it having been ejected from
   R136, and this star therefore sets a robust lower limit on the age of the central
   cluster of $\sim$1.3 Myr.
   }


   \maketitle
   
%

\section{Introduction}

The presence of
very massive (M$\gtrsim$100\,M$_{\sun}$) isolated stars within $\sim100$ parsec
of extremely young massive clusters such as R136 and Westerlund 2
\citep{walborn2002, evans2010, romanlopes} is interpreted as evidence that the
dynamical ejection scenario (DES; \citealt{poveda}) is an effective mechanism 
for ejecting some of the most
massive stars from their birthplaces.  The competing channel of ejection by the
binary supernova ejection scenario (BES; \citealt{blaauw}) may be excluded
since, with ages of less than
about 2 Myr \citep{sabbi2012, crowther2016, zeidler}, 
these clusters are unlikely to have produced a core-collapse SN,
an event even less likely when these stars were ejected.
While it has been suggested that some
isolated massive stars may form from small molecular clouds 
\citep{parker, bressert, lamb2016}, the peculiar line of sight ($LOS$) velocities of
some of the most massive stars indicate that they are 
runaway stars and strong candidates for dynamical ejection.
In fact, N-body simulations of dynamical ejection from star clusters
\citep{banerjee2012,perets, ohkroupa} have had some success in
explaining the presence of the isolated very massive stars around 
clusters like R136, as has the model in which a 
single very massive wide binary, a "bully binary",
scatters massive stars out of their host cluster after a single interaction \citep{fujii}.

The proto-typical example of such an object, presented by \cite{evans2010}, is the star VFTS~16,
a $\sim$100\,M$_{\sun}$ O2\,III star on the periphery of 
30 Doradus within the Tarantula Nebula of the Large Magellanic Cloud. 
This star is $\sim$120\,pc in projection from the central cluster R136 and
data from the VLT-FLAMES Tarantula Survey \citep[VFTS,][]{evans2011} 
shows that it has a LOS radial velocity
which is discrepant by $-85$\,$\kms$ compared to the central cluster, R136, while
multiplicity is ruled out by the survey's multi-epoch data. 
In the absence of any measured 
proper motion for VFTS~16, \cite{evans2010} proposed
R136 as the parent cluster based on the star's
main-sequence lifetime, assuming a transverse velocity comparable to its peculiar
radial velocity, and given the youth and mass of R136. 
They further argued that VFTS~16 may be one of the
clearest cases for ejection of a very massive star by dynamical interaction
from a young massive cluster, with \cite{banerjee2012} presenting an N-body
simulation in support of this idea.

In this paper we present the {\em Gaia} DR2 proper motion for
VFTS16, and demonstrate that its magnitude and direction are fully
consistent with R136 as its point of origin, reinforcing the idea that the star has indeed
been ejected from the cluster.
In section 2 we present the {\em Gaia} data and
analysis details upon which our conclusions are based, while 
in section 3 we discuss their potential significance.

\section{Analysis of {\em Gaia} DR2 data}

\subsection{Defining the local reference frame of R136}

Using the $Gaia$ Data Release 2 (DR2) 
catalog \citep{gaiadr2a} we extracted all sources within 0.2 degrees of R136 
after first screening out stars fainter than $G=17.0^m$, 
corresponding to the approximate faint limit of the VFTS (given
the significant extinction of OB stars in
this field we have that $G\approx V$).  This sample was cross-matched
with VFTS sources and other known massive stars before
filtering out foreground stars using their parallax measurements,
and measured radial velocities (also from VFTS).

We noticed that the resulting sample of 827 sources
had considerable numbers of stars with very high tangential velocities
relative to their surroundings, up to 250 $\kms$ in some cases.  
These high velocities are clearly spurious as we know
from spectroscopy that the $LOS$ relative velocities should be less than
$\sim$100 $\kms$. Dividing our sample into {\em fast} (202 sources, or 24\% of the
filtered sample) and {\em slow} (625 sources) subgroups,
with the divide being arbitrarily chosen to be a tangential velocity of 100 $\kms$, we noted a clear 
difference in distributions with the {\em fast} subgroup being more strongly
concentrated in areas of high stellar density and/or strong nebulosity, such as in
the centre of the field near R136 itself (i.e. NGC2070),
as shown in Figure 1. This distribution, with the fastest moving stars
more strongly clumped, is also the opposite of what one expects if
these proper motions were real. We also see some apparently fast moving
stars in more isolated environments, however spot checking these 
it appears they are often either blended sources (OB stars) or
late-type LMC field stars. 

As discussed by \cite{gaiadr2b} the presence of close companions 
may contribute to delivering spurious astrometric solutions and indeed we confirmed
that in some cases, using existing multi-colour $HST$ imaging of our field 
from the Hubble Tarantula Treasury Project (HTTP; \citealt{sabbi2013}), 
stars with very high proper motions were indeed blended sources.
While \cite{gaiadr2b} discuss potential filtering approaches,
we found that most of the sources in our sample with spuriously high proper motions  were 
removed by excluding those
sources with proper motion errors greater than $0.1\masyr$
in both co-ordinates. Filtering our original $Gaia$ sample with this constraint
produced a subsample of 682 sources. This subsample was
found to have mean proper motion components 
of pmRA=1.72 and pmDec=0.67$\masyr$. Restricting the sample to the
central 153 stars within 0.05 degrees of R136 (thus including NGC2060), 
on the other hand, resulted in
only slightly different values of pmRA=1.74 and pmDec=0.70$\masyr$, with
standard deviations of 0.13 and $0.20\masyr$ respectively (the error in the means
being $\sqrt{N}$ smaller; 0.01 and $0.02\masyr$ respectively).
Since we are primarily interested in testing the hypothesis that VFTS16 was
ejected from the central R136 cluster, we have converted 
all absolute proper motions to proper motions relative to this region by subtracting
the mean motion of the central stars from the sample.

   \begin{figure*}
   \includegraphics[scale=0.3]{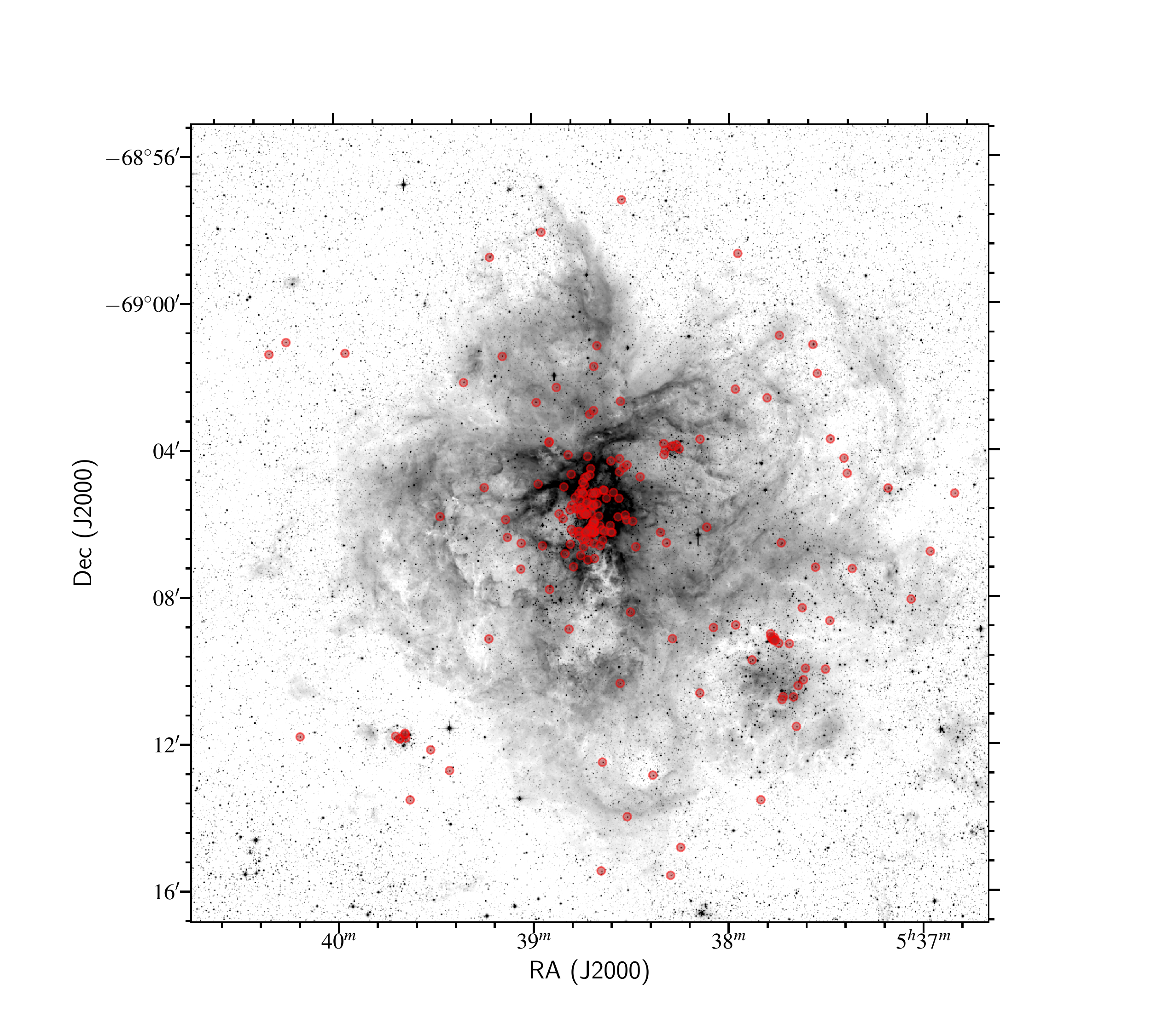}
%
   \includegraphics[scale=0.3]{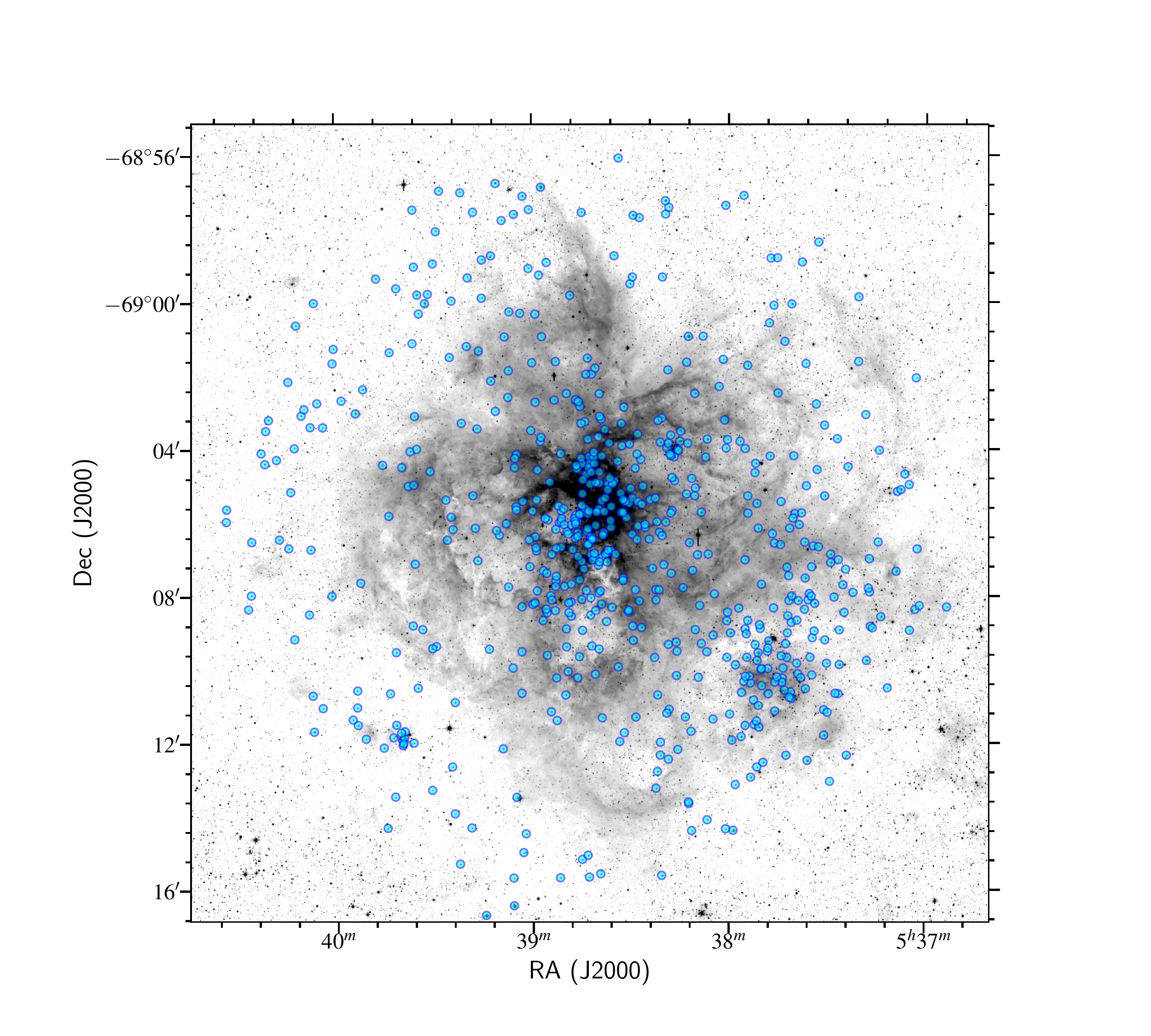}
   \caption{On the left (red points) we have the {\em fast} sample of stars (202 sources) with assumed spurious tangential velocities 
   greater than 100 $\kms$, while on the right (blue points) we illustrate the {\em slow} sample of stars
   (625 sources) with tangential velocities less than 100 $\kms$. The {\em fast} sample, 24\% of the total, are
   clearly clumped within regions of high stellar density and/or strong nebulosity.}
              \label{Fig1}%
    \end{figure*}
    
\subsection{The relative proper motions of the O-type stars}

We cross-matched the resulting catalog against the O-type stars
taken from the VFTS catalog \citep{evans2011} 
and obtained matches for 193 sources. However it was found that the proper motion
diagram of this sample still had a significant number of outliers, therefore
we applied a further, stricter, filter by excluding all sources
with proper motions errors greater than 0.07$\masyr$ in both ordinates,
resulting in a final selection of 79 sources.  While this procedure
undoubtedly removes many bona fide measurements, our primary
objective is to set the context for VFTS~16, with errors of approximately 
0.05$\masyr$, and for that purpose
the procedure is sufficient. This is illustrated in Fig.~2
where one can see that due to the rapid increase in the dispersion of
the proper motion measurements for stars with
errors greater than $\sim0.07\masyr$, it would be impossible to 
unambiguously distinguish between genuine runaway candidates
and outliers if we include these stars
in a proper motion diagram. Moreover, from Fig.~B.2 of 
\cite{gaiadr2a},  0.07$\masyr$ is the median
value of the formal uncertainty of the proper motion for all
sources in the Gaia
catalog at the approximate median magnitude of our O-star sample 
($G\sim15.3$). 


   \begin{figure}
   \includegraphics[scale=0.5]{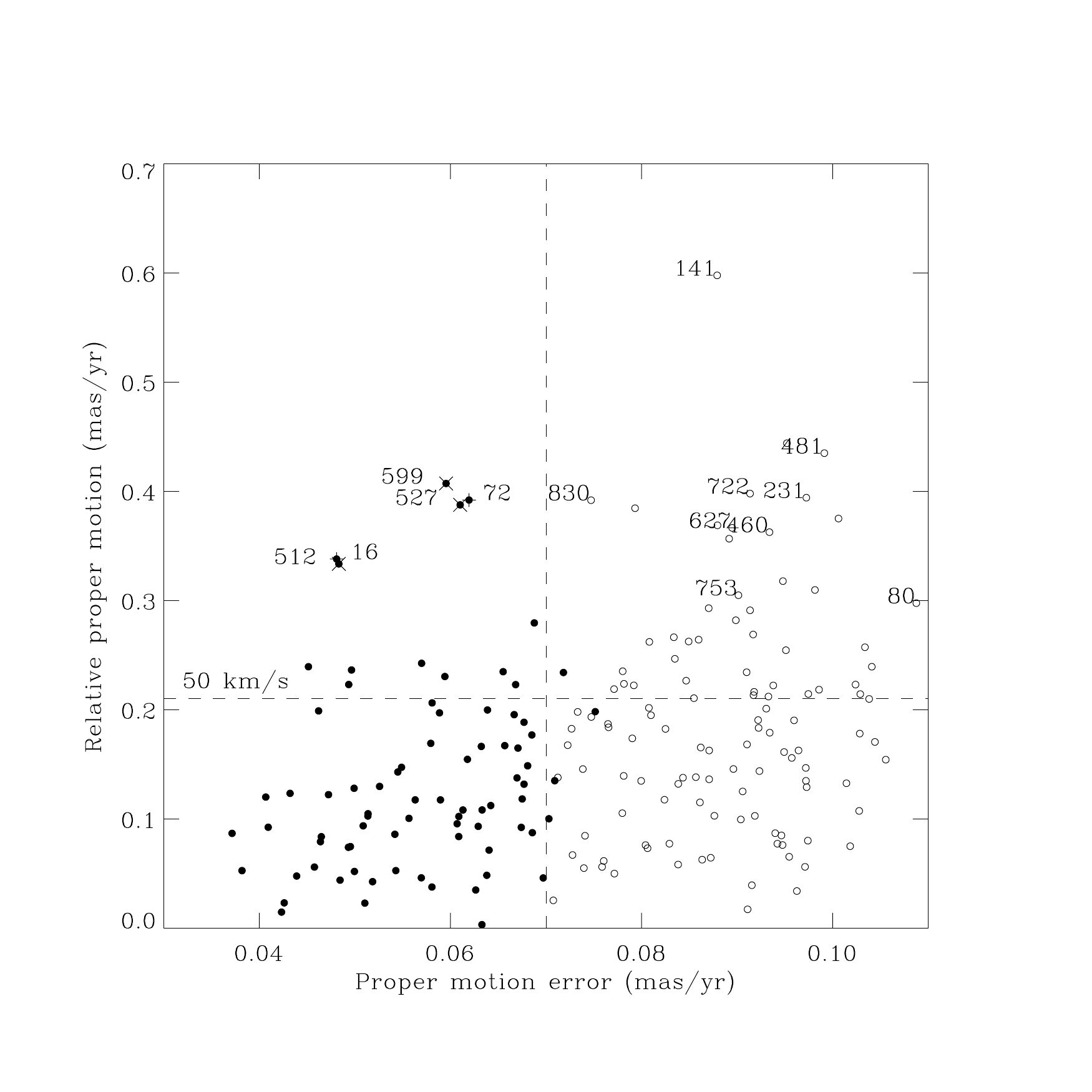}
%
   \caption{Comparison of relative proper motion versus the magnitude of the semi-major
   axis of the error ellipse of the proper motion 
   for all 193 O-stars in our core sample, with the filled symbols
   representing those stars with errors less than 0.07$\masyr$ in both proper motion 
   ordinates. Ignoring 
   outliers, it is apparent that the dispersion of proper motions begins to increase 
   significantly above 0.07$\masyr$ (vertical dashed line). For illustration the horizontal 
   dashed line represents a velocity of 50$\kms$ at the distance of the LMC. Labels are
   the VFTS identification numbers for those stars discussed in the text.  
   }
              \label{newfig}%
    \end{figure}

In  Fig.~3a we display the resulting proper motion diagram,
where one can see that we now
have only a small number (5) of clear outliers, with relative 
proper motions in excess of $0.3\masyr$.
The distribution of the bulk of the stars appears to be 
offset to negative velocities with respect to the origin (i.e. R136).
However checking the distributions of these stars it is clear that
this is due to the large number of O-type stars
in NGC2060 (see Fig.~4) that have on average a bulk velocity
with respect to R136/NGC2060 of approximately $-0.06\masyr$ in RA, or
15 $\kms$, and about $+0.02\masyr$ in DEC, or about 5 $\kms$.

We made a final manual
review of outliers in this diagram to check for crowded stars, or stars 
contaminated with strong nebular emission. Three of these objects; 
the O2\,III-V((f*) spectroscopic binary VFTS~512 (and an X-ray source), 
the O6.5\,Iafc+O6\,Iaf binary VFTS~527 (alias R\,139 and also an X-ray source) 
and the O2\,III(f*) star VFTS~599, are in the crowded
inner region of NGC\,2070 where we detect many spuriously large proper motions.
We suspect some, or all, of these three measurements may not be reliable and
defer discussion of these objects pending further analysis of the 
impact of crowding and/or strong nebulosity on the $Gaia$
astrometry.

In contrast we confirm that VFTS16 is isolated,
and resides in a region of very low stellar density, consistent with
its runaway status from \cite{evans2010}. 
Its $Gaia$ parameters indicate that the source was well observed (having {\tt visibility\_periods\_used}=17), indicating high astrometric quality with very low DR2 uncertainties. This is also confirmed by the {\tt astrometric\_sigma5d\_max} (being 0.055 mas) which represents the semi-major axis of the position error ellipse and is therefore useful for filtering out cases where one of the five parameters, or some linear combination of several parameters, was particularly ill-determined. (Gaia Collaboration et al. 2018a).

VFTS~72 is also an obvious outlier and we confirm that
this is also isolated in the $HST$ images. As discussed by \cite{evans2010}
this O2 III-V(n)((f*)) star was also suggested as a potential 
runaway star in 30 Doradus \citep{walborn2002},
however its $LOS$ velocity of 273 $\kms$ is both very close to the mean O-star
velocity of $~$270 $\kms$, and shows no significant variation \citep{sana2013}. We therefore confirm 
it as a runaway based on its proper motion with respect to R136. 
Similarly to VFTS~16, its $Gaia$ DR2 parameters indicate that VFTS 72 has an
astrometric quality that is very high (a {\tt visibility\_periods\_used} of 16 and  an
{\tt astrometric\_sigma5d\_max} of 0.067).

For each star the PM case for runaway status is statistically significant. The joint 
probability of a PM as large as observed (or larger) and an angular alignment as 
well as observed (or better) jointly occurring by chance, for a PM distribution
with the RMS values in Figure 3a, is less than 1\%. Given that we have studied a 
sample of 79 stars in Figure 3, the probability a finding a single 1\%-unlikely PM event is
non-negligible. However, the fact that for both these stars there is independent 
evidence for runaway status already in the literature, as discussed in this paper,
strongly suggests that these are not just chance alignments in the normal tail of the PM
distribution.

 \cite{platais2018} have also presented proper motions for
stars in 30 Doradus and the Tarantula Nebula based on two epochs of $HST$
observations.
However VFTS~16 was not included in that catalogue because the star
is saturated in all exposures, and VFTS~72 was outside the field covered by the
catalogue (though the star was observed in some exposures of the
subsequent HTTP survey). 
However a new method to determine the proper motions
of saturated stars in $HST$ images (Anderson 2018, priv. comm.) does suggest a 
proper motion for VFTS~16 that is consistent with the $Gaia$ DR2 measurement, lending added confidence
to the main result of the present paper.

For visualisation, we also decomposed the proper motions of the sample
stars into radial and tangential proper motions relative to R136.
Figure 3b shows the resulting scatter diagram. A star that is moving
rapidly and purely radially away from R136 would appear near the x-axis 
to the far right of the plot. Indeed, VFTS~16 and VFTS~72 are found in this
area, confirming their status as candidate runaway stars. We do note 
that one of the other rapidly moving stars, VFTS~512, also has a nearly 
radially directed motion.
Therefore its measured DR2 proper motion could possibly be correct, and indicative of runaway status. 
When we relax our conservative sample cut on the proper motion errors of 0.07$\masyr$,
several more stars appear near VFTS ~16 and VFTS~72 in our polar plot of proper motions, namely
VFTS~80, 460, 481, 627, 722, 753 and  VFTS~231, 830 respectively.
However from Fig.~2 one can see that their nature is somewhat ambiguous and,
in the absence of a better understanding of the DR2 systematic uncertainties discussed
in section 2.1 (and illustrated in Fig.~1), we
refrain from interpreting the motions of these stars here. 
 For example our most extreme outlier in proper motion from Fig.~2
is VFTS~141, however close inspection of HST images of this star
confirm it to be a very close visual binary and hence we 
consider this measurement unreliable.
Nevertheless, it is possible that with
additional analyses or future Gaia data releases, some of them
may prove to be true runaways. Hereafter, we focus exclusively on the cases of 
VFTS~16 and VFTS~72.

Properties for both of these stars are collected in Table 1, 
where we
convert proper motion to velocity assuming a distance to
the LMC of $50.1\pm2.5$ kpc, corresponding to a
distance modulus  of m-M=$18.50\pm0.1$ \citep{freedman}.
We note that the distance uncertainty also implies a
$\sim$5\% systematic uncertainty in predicted and derived velocities. 

\begin{table*}
\begin{centering}
\caption{Properties of the candidate runaway O2 stars. Relevant stellar parameters
in the upper section of the table are from the VFTS as indicated, while in the lower
section we give the $Gaia$ source identifier, relative proper motion components, 
position angle of proper motion (PA$_{\rm pm}$), distance in 
projection to R136 ($d_{\rm R136}$) in arcsec, and position angle (PA$_{\rm R136}$) of source with
respect to R136 (East of North) in degrees,  
and time of flight $t_{\rm R136}$ from R136, calculated simply as $d_{\rm R136}$/(rel. pm).
}
\begin{tabular}{lll}
\hline\hline
 & VFTS~16 & VFTS~72 \\
 \hline
 \\
 Spectral Type\tablefootmark{a} & O2\,III & O2\,III-V(n)((f*) \\
 T$_{\rm eff}$\tablefootmark{b}  (K) & 50\,600$^{+500}_{-590}$& 54\,000 $\pm1500$\\
 Current Mass\tablefootmark{b}  (M$_{\odot}$) & 91.6$^{+11.5}_{-10.5}$& 97.6$^{+22.2}_{-23.1}$\\
  Age\tablefootmark{b} (Myr)  & $0.7\pm0.1$ & $0.4^{+0.8}_{-0.4}$ \\
  $v$sin$i$\tablefootmark{c} ($\kms$) & 112$\pm30$& 185$\pm30$ \\
  $v_{LOS}$\tablefootmark{d} ($\kms$) & 189.2$\pm1.2$ & 273.4$\pm1.9$\\
  \\
$Gaia$ Source id & 4657690620070706432 & 4657698454092124416  \\
rel. pmRA ($\masyr$) & $-0.336\pm0.046$ & $-0.372\pm0.050$ \\
rel. pmDec ($\masyr$) & $-0.038\pm0.045$& $+0.125\pm0.061$ \\
  PA$_{\rm pm}$  & $-96^{\circ}$$\pm8$ & $-71^{\circ}$$\pm8$ \\
 PA$_{\rm R136}$ & $-99^{\circ}$ & $-51^{\circ}$ \\
  $d_{\rm R136}$ (arcsec) & 506 & 370 \\
rel. pm ($\masyr$) & $0.338\pm0.048$ & $0.392\pm0.062$ \\
tangential speed ($\kms$)\tablefootmark{e} & $80\pm11$ & $93\pm15$ \\
3D speed\tablefootmark{f} ($\kms$) & $112\pm8$ & $93\pm15$ \\
 $t_{\rm R136}$  (Myr)& $1.50\pm0.21$ & $0.94\pm0.15$ \\
 \\
 \hline
\end{tabular}
\tablefoot{
\tablefoottext{a}{\cite{walborn2014}}
\tablefoottext{b}{\cite{schneider2018} but see the Sect.~3}
\tablefoottext{c}{\cite{oscar2013}}
\tablefoottext{d}{\cite{sana2013}}
\tablefoottext{e}{Assuming a distance to the LMC of 50.1 kpc}
\tablefoottext{f}{We adopt 267.7 $\kms$ as the $LOS$ velocity of R136 from \cite{vftsVII}}
}
\end{centering}
\end{table*}

   \begin{figure*}
\centering
   \includegraphics[scale=0.65]{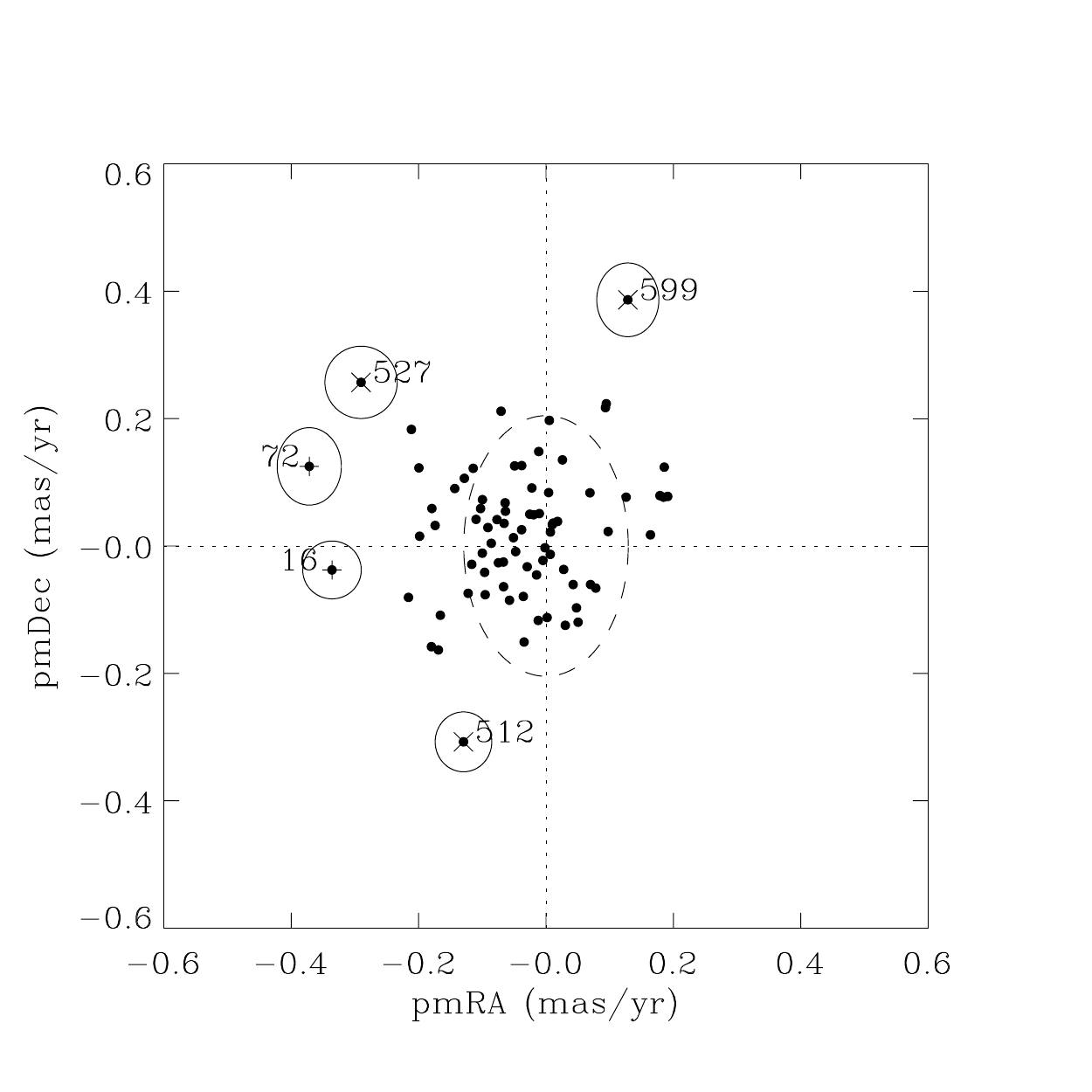}
   \includegraphics[scale=0.65]{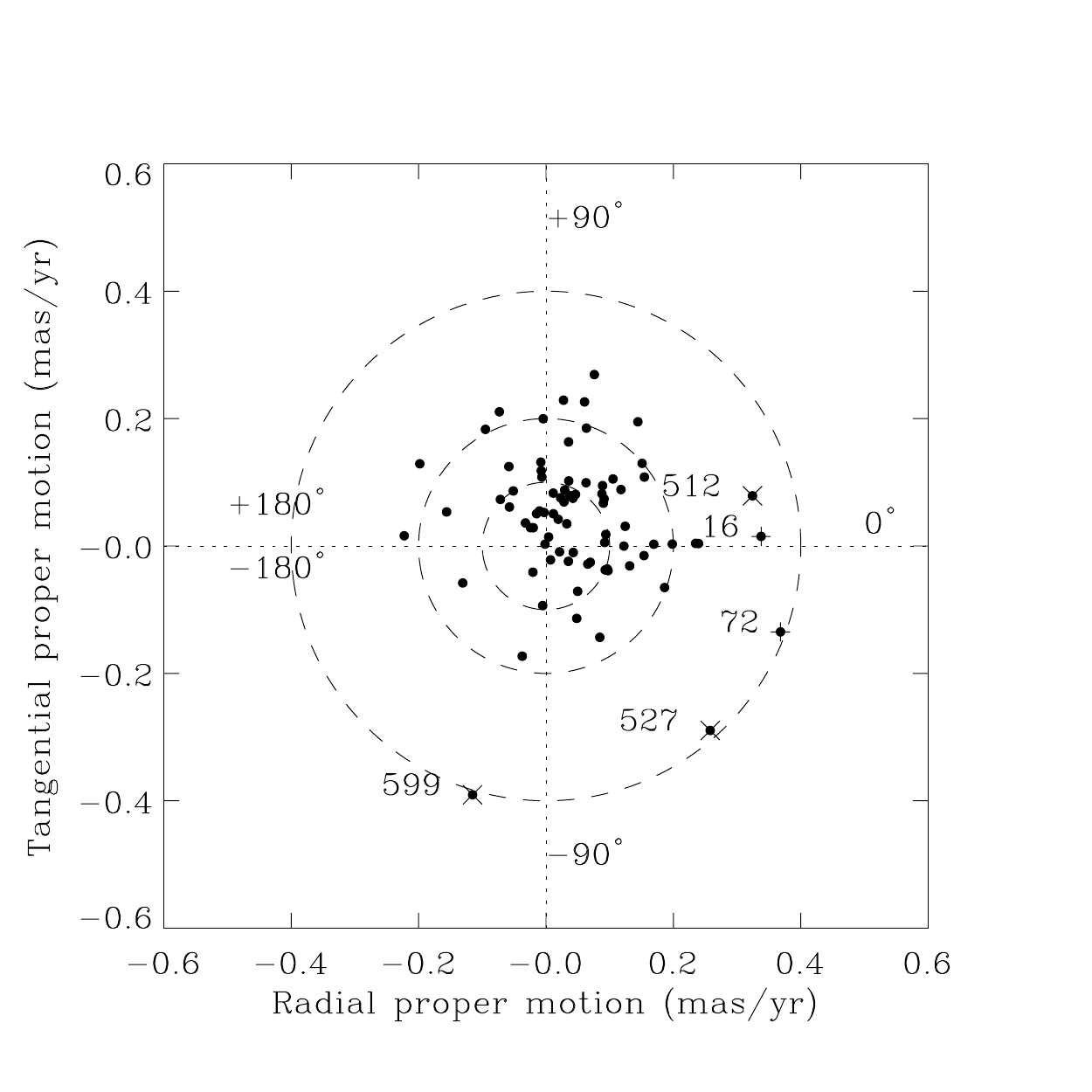}
   \caption{Relative proper motions (0.1$\masyr$$\approx$25 $\kms$) for the 79
   O-type VFTS stars in 30 Doradus that have proper motion errors less than 
   0.07$\masyr$. VFTS~16 and 72 are labelled (marked with +), as are three additional outliers
   (marked with $\times$) that are
   discussed briefly in the text. The left-hand figure {\bf (a)} is the standard proper motion diagram with
   the error ellipses indicated for these 5 stars, though for clarity
   the ellipses for the full sample are omitted. The inner dashed ellipse denotes the velocity
   dispersion of the local reference frame around R136/NGC2070 sample discussed in section 2.2.
   In the right-hand figure {\bf (b)} we show a polar plot of the radial and 
   tangential components relative to
   the direction from R136, concentric circles denoting relative proper motions of 0.1, 0.2 and
   0.4$\masyr$. In this diagram the positive x-axis, $0\degr$, denotes a direction radially outward from R136,
   with positive angles indicating that the tangential component is counterclockwise with respect
   to the position angle of the star relative to R136.
    }
              \label{Fig2}
    \end{figure*}

\section{Discussion}

 \cite{evans2010} argued that VFTS~16 was
a candidate for dynamical ejection from R136 based on 
various of pieces of circumstantial evidence; it is a $LOS$ runaway and
it is very massive and so young that R136 is the only potential
launch site, even when accounting for plausible rejuvenation of the 
runaway star, provided its high peculiar $LOS$ velocity ($-78$ $\kms$) would be
matched by its tangential velocity.  The case for VFTS~72 being a runaway 
was first put forward by \cite{walborn2002} and
was based primarily on the discussion of this
star's relative isolation in the field near 30 Doradus,
albeit from VFTS we have that this star's radial velocity
is not anomalous \citep{sana2013}, and exhibits little sign of
variability.  In the above we have determined that VFTS~16 and VFTS~72 are
indeed runaway stars in the tangential plane, with relative tangential
speeds of 80 and 93 $\kms$ respectively. 

In the following we discuss the consistency of the various
timescales of interest for each star, namely, their flight times to R136, 
their ages (for now assuming single star evolution), and the age of
the central R136 cluster. For the age of the cluster we adopt the
value proposed by \cite{crowther2016} of $1.5^{+0.3}_{-0.7}$ Myr
as this is based on $HST$/STIS spectroscopy of the most massive
stars within the central parsec of the cluster.

   \begin{figure}
\centering
   \includegraphics[scale=0.4]{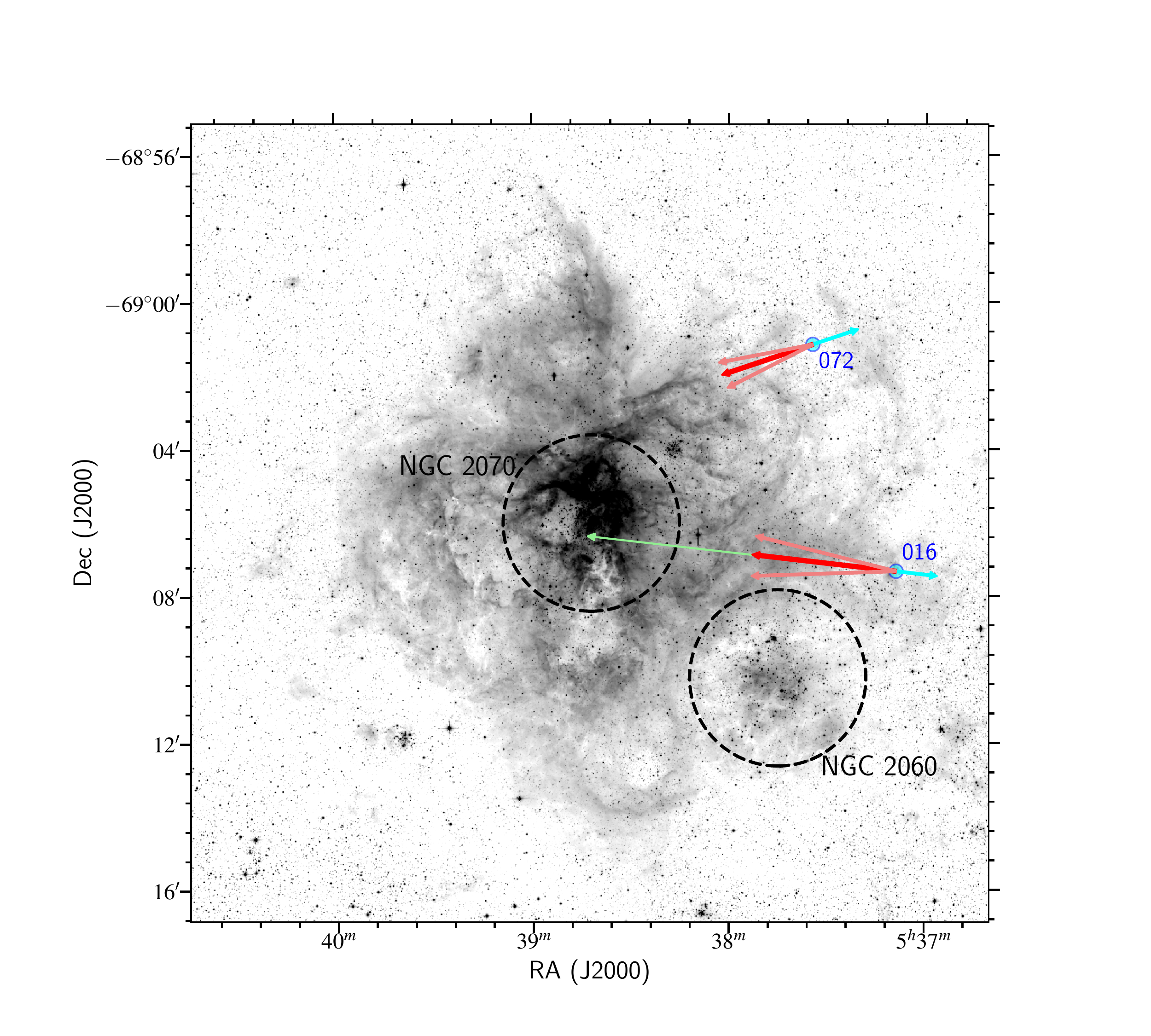}
   \caption{The proper motions of VFTS~16 and 72
   relative to the cluster R136,
   that resides within the centre of the association NGC2070. The lengths of the 
   arrows are scaled to their tangential proper motions.  Red arrows indicate the 
   projections of these stars' proper motions backwards in time, scaled to their ages,
   and their approximate opening angles, from Table 1. The green arrow denotes the
   denotes the distance the star would cover to R136. The other major
   grouping of O-type stars in this region, NGC2060, is also indicated.  
   }
              \label{Fig3}
    \end{figure}

In Table 1 we present the flight time to R136, $t_{\rm R136}$, 
simply dividing the angular distance to the cluster, $d_{\rm R136}$, by
the tangential proper motion. This is of course only correct if the proper motions
align exactly along the position angle to the central cluster, which is a good
approximation for VFTS~16 but not for VFTS~72, as
shown in Figure 4, and in Table 1. In addition these values do not take account
of possible deceleration as the star leaves the cluster, and as such should be
interpreted as upper limits (though the effect is small).
Nevertheless, considering VFTS~16 we find that 
the position angle of its velocity vector with respect to R136 is $\sim +3\degr$,
with an opening angle of $\sim17\degr$, which is consistent
within the uncertainties. Also from Table 1 we have $t_{\rm R136}\sim$1.5 Myr, 
to be compared with the age 
estimate provided by \cite{schneider2018} of  $0.7\pm0.1$ Myr. Clearly
the latter value is in tension with the flight time though we will return to
this issue below. For now we note that $t_{\rm R136}$ is in good agreement
with the age of R136 of $1.5^{+0.3}_{-0.7}$ Myr, also noting
that this timing implies that ejection would have to have occurred very
early in the lifetime of the cluster. The case of VFTS~72 
is a little different. Its position angle is 20$^\circ$ from alignment with
the central cluster,  a discrepancy of $\sim$2$\sigma$.
\cite{schneider2018} determined the age of VFTS~72 as  $0.4^{+0.8}_{-0.4}$ Myr,
somewhat smaller than $t_{\rm R136}$, but consistent within the uncertainties.

\cite{fujii} presented a DES model, the "bully binary" model,  for the ejection of massive 
stars from dense young clusters, focusing on R136 as an important test case and in fact
proposing that VFTS~16 is one of the ejected stars predicted by
their model.  In this model,  a very massive wide binary that formed during
the collapse of a young star cluster acts as a scattering source due
to its high collision cross section. The scattered stars may attain velocities
well in excess of the cluster escape velocity, and \cite{fujii} predict that a 
cluster like R136 should have produced $\sim$5 runaways with velocities
greater than 30 $\kms$, and with masses greater than 8 M$_{\odot}$. While
there are promising properties of the bully binary model to explain 
runaways around massive clusters, a potential problem with this model
is that it has difficulty producing very massive runaways with space velocities as
high as VFTS~16 (112 $\kms$) because in this model it is assumed the
runaway is ejected after a single interaction with the bully binary.
Models that allow for multiple interactions with multiple binaries 
do predict faster runaways \citep{ohkroupa,perets} though the fraction
of runaways produced with velocities greater than $\sim$100 $\kms$
is typically very small, a few percent or less. In a similar N-body simulation 
taylored specifically to R136, \cite{banerjee2012} demonstrate that two of their
four simulations of R136 each produce a single runaway with properties similar to
VFTS~16 within the first Myr. If VFTS~72 would also be a DES runaway, that would
begin to strain the predicted numbers of such very massive and fast runaways
by roughly a factor of 2--4. (See also Renzo et al., to be submitted, for a further 
candidate.) However given the caveats expressed in section 2.2
that will affect the completeness of our sample, in particular the difficulty
in determining proper motions of stars close to R136, it is not useful
at the present time to make statistical comparisons.

As mentioned above, the inferred age of VFTS~16 
($0.7\pm0.1$ Myr) is significantly less that its flight time ($1.50\pm0.21$ Myr)
and also just consistent with the age of R136 ($1.5^{+0.3}_{-0.7}$ Myr) within the
errors.  It is tempting to argue that VFTS~16 star might be a merger
product of an ejected binary (see \cite{oh2014} for a simulation of how
the very massive binary R144 may be a DES runaway from R136). 
Indeed stellar mergers may lead to rejuvenated stellar products \citep{schneider2016, demink2014}
and thus provide a potential channel to explain the age versus flight time discrepancy.
However VFTS~16
has a rather low $v$sini of 112 $\kms$, and a surface nitrogen abundance 
that implies an initial rotational velocity not much larger than that (see Figure 5).
While its properties do not naturally suggest that it is a
merger product, it is worth recalling that the massive runaway stars predicted 
by \cite{banerjee2012} are, in fact, mergers. 

The ages in Table 1 are from \cite{schneider2018} and are based on the FASTWIND analysis of
the VFTS optical data from \cite{oscar} and \cite{sabin2017} for VFTS~16 and VFTS~72
respectively, and make use of the evolutionary tracks of \cite{kohler} for
LMC metallicity.
In the case of VFTS~16  the small uncertainty in the age is
driven primarily by the adoption of a $1\sigma$ formal error 
(68\% confidence limit) in the
effective temperature,  T$_{\rm eff}$=50\,600$^{+500}_{-590}$K. We note that \cite{oscar} 
quote the 95\% confidence intervals of
$^{+500}_{-1190}$K in T$_{\rm eff}$ for this star, while \cite{bestenlehner2014}
derived a higher T$_{\rm eff}$ for VFTS~16 of 53\,100\,K using CMFGEN,
quoting an error of 0.02 dex, or approximately 2\,500\,K. 
Analogous comments apply to VFTS~72. It has been analysed a number of 
times using FASTWIND  and CMFGEN with T$_{\rm eff}$ values ranging
from 50\,000\,K to 55\,000\,K \citep{mokiem2007, doran2011, rivero2012, bestenlehner2014,
sabin2017}. 

To examine the sensitivity of the age of VFTS~16 to systematics 
in the determination of $T_{\rm eff}$ we
assumed a larger uncertainty of $^{+2000}_{-1000}$\,K in the effective 
temperature and,
using {\sc bonnsai}\footnote{The BONNSAI web-service is available at www.astro.uni-bonn.de/stars/bonnsai} \citep{bonnsai,schneider2017}, we derive an age of $0.9^{+0.3}_{-0.2}$ Myr.
It is difficult to argue for an age greater than this as the upper 
bound is quite robust.
As discussed by \cite{walborn2002}, in their paper defining the 
O2 spectral type, these stars are characterised primarily 
by the lack of He\,{\sc i} lines in the optical. Consequently the
lower bound on T$_{\rm eff}$, and the upper bound on the age, is provided 
by the presence of a weak He\,{\sc i} 4471 line in the models, and its absence
in the observations.  Lowering the temperature below 50\,000-49\,000\,K
gives rise to He\,{\sc i} in the models that is not detected in the data.
Turning to VFTS~72 we consider T$_{\rm eff}$=54\,000$^{+1000}_{-4000}$K
as a compromise between published parameters.  
We also adopt a nitrogen abundance of 8.2$\pm0.3$ dex derived from 
the same VFTS data and methods described 
in \cite{sabin2017} , and with these modified parameters we derive an 
age of $0.80^{+0.35}_{-0.47}$ Myrs, consistent with the results of
\cite{schneider2018}. We illustrate the sensitivity of the ages to
the stellar parameters and evolutionary tracks in Fig.~5 
(the nitrogen abundance for VFTS~72 is from \citealt{grin}).
Summarising this part of the discussion, even allowing for the range of
published effective temperatures of these stars,  
it is difficult to reconcile the age of VFTS~16 with the flight time
of the star from R136. 


In order to assess the sensitivity of the age discrepancy on the
underlying stellar evolution models, we reviewed the literature and
found that a 100\,M$_{\odot}$ model including rotation for half-solar metallicity
is provided by \cite{choi}. However, their model
rotates initially very fast ($v_{\rm rot,initial}/v_{\rm crit}=0.4$,
or $v_{\rm rot,initial} \simeq 450\kms$), and reaches CNO-equilibrium
abundances at the surface very quickly, evolving from the ZAMS
($T_{\rm eff}\simeq 54\,000\,$K) only to hotter surface temperatures
(solid red track in Fig.~5).
The use of solar scaled initial abundances by \cite{choi} limits the surface
nitrogen enrichment to a factor of order\,10, while both observations and
models using LMC initial abundances suggest consistently a maximum
enrichment factor of the order of 30 in LMC massive stars \citep{hunter09}. 
To mimic this, we have doubled the N-enrichment
in the track of Choi et al.\, (dotted red track in Fig.~5). Using this enhanced
abundance, their model reaches the level of nitrogen
enhancement as observed in VFTS16 after about 0.3\,Myr. At an age
of 1.5\,Myr, its surface has reached CNO equilibrium, and the model has
left the plot area to the left in Fig.~5.

Thus, in agreement with the results from \cite{kohler}, models of fast rotators can not represent 
VFTS16. This is confirmed by the rotating ($v_{\rm rot,initial}/v_{\rm crit}=0.4$) LMC models shown by 
\cite{crowther2010}. While nitrogen enhancement is not discussed in detail in their 
paper, at an age of 1.5\,Myr all their models are strongly helium-enriched, which implies CNO-
equilibrium surface abundances \citep{grin}.

Thus, the interpretation of VFTS~16 in terms of single star models requires moderate initial rotational 
velocities, in the range $v_{\rm rot,initial}\simeq 150\dots 200\,$km\,s$^{-1}$. Except for those of 
\cite{kohler}, corresponding models are not available in the literature. However, as studied by 
\cite{sanyal}, the effective temperature evolution of the models by \cite{kohler} is 
affected by envelope inflation, as they are very close to the Eddington limit. They showed that 
the degree of envelope inflation, and thus the effective temperature at a given time, depends on the 
efficiency of energy transport in the subsurface convection zones of these models. At an age of 
1.5\,Myr, the radii of the models of K\"ohler et al. shown in Fig.\,5 are inflated by about 20\%, implying 
that corresponding non-inflated models would be about 5000\,K hotter. While there is empirical and 
theoretical evidence for the inflation phenomenon 
\citep{petrovic,grafener}, its 
extent is uncertain. The models of \cite{kohler} could therefore underestimate the effective 
temperatures, and we can not exclude the possibility that VFTS~16 is an evolved single star with an 
age of 1.5\,Myr.

Nor can we exclude the possibility that either star might be a BES runaway, though clearly this is 
more relevant to VFTS\,72 whose proper motion direction is approximately 2$\sigma$ away from its 
position angle with respect to R136. Whereas population synthesis studies mostly predict smaller 
kinematic velocities then those we derived for our two stars \citep{eldridge,renzo}, the fastest BES 
runaways may come from binaries which undergo common envelope evolution (see for example 
\cite{boubert2017} which is essentially unconstrained at very high mass. In any case, the age 
problem of both stars is even more severe in the BES scenario, since they would necessarily have to 
be older than the shortest stellar life time of about 2\,Myr \citep{kohler}. This remains true even 
when considering a potential rejuvenation of a BES runaway due to mass accretion, since the stellar 
life times at very high mass depend only weakly on the stellar mass. We therefore consider a BES 
origin for both stars as unlikely, most strongly so for VFTS\,16, for which also the weak nitrogen 
enrichment argues against an accretion history. 

   \begin{figure}
\centering
   \includegraphics[scale=0.3]{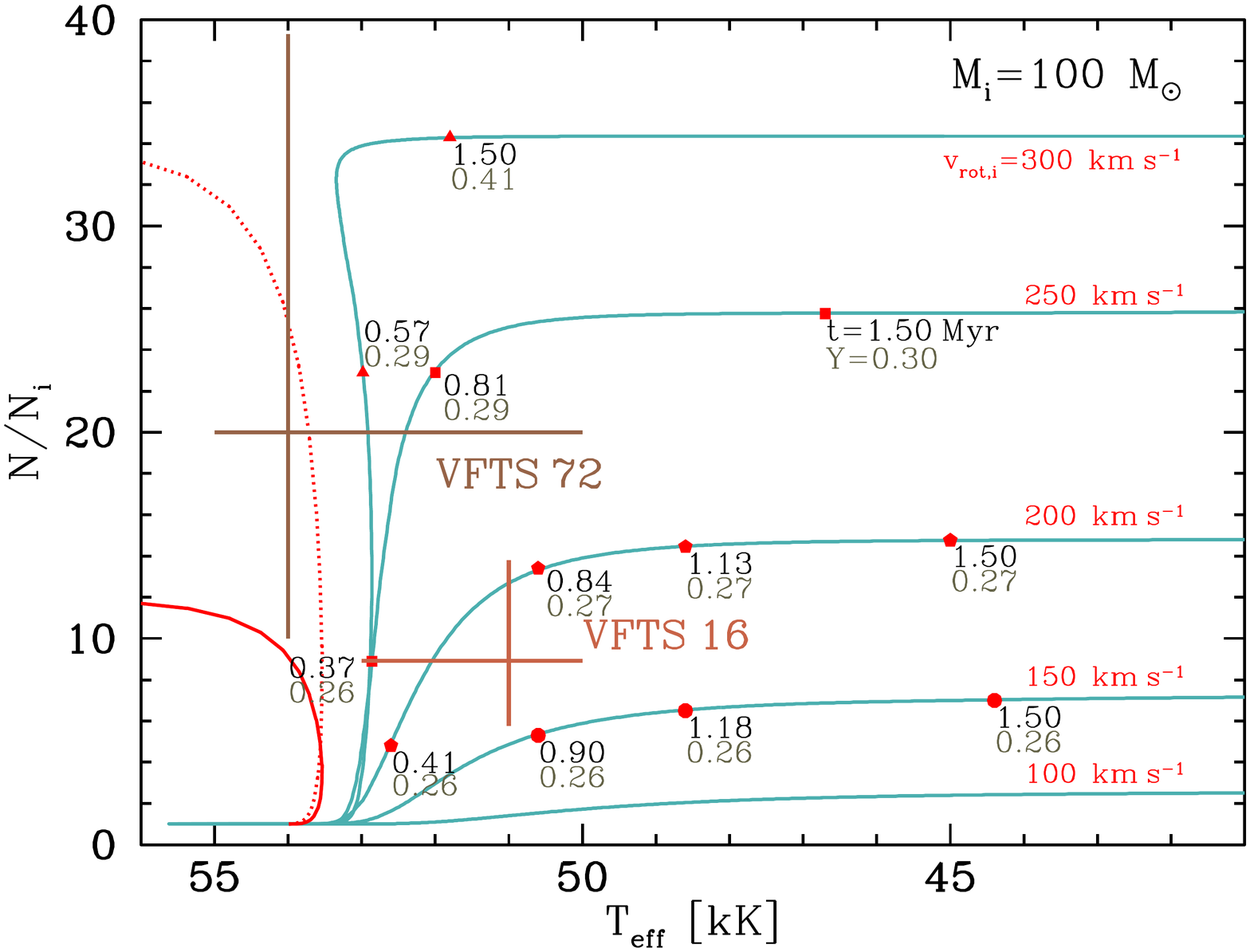}
   \caption{In this figure we illustrate the positions of VFTS~16 and 72 in the 
   surface nitrogen enrichment versus effective temperature diagram, superimposed
   on the evolutionary tracks of \cite{kohler} for LMC metallicity (in blue). 
   The tracks are labelled with their initial 
rotational velocity, with time-steps indicated along with surface helium
abundance. Since O2 dwarfs and giants typically have
$T_{\rm eff}\gtrsim 50\,000$\,K  one can see that for
such a star to have an age of $\sim$1.5 Myr it would be
significantly more nitrogen rich than is derived for VFTS~16, and it should be
helium rich (also not inferred from the observations). For comparison we also 
show tracks (in red) from \cite{choi}, as discussed in section 3.} 
              \label{Fig4}
    \end{figure}

\section{Conclusions}

Consideration of the $Gaia$ DR2 proper motions of O-type stars in
the region of the Tarantula Nebula provides strong support for the
hypothesis proposed by \cite{evans2010} that the $\sim$100\,M$_{\odot}$
runaway star VFTS~16 was ejected from the central cluster R136
by the mechanism of dynamical interaction with extremely massive
binaries in the cluster. 

We have also discovered that another 
isolated $\sim$100\,M$_{\odot}$ star
in the region, VFTS~72, is also a proper motion runaway, as
previously suspected by \cite{walborn2002} on the basis of its relative 
isolation (it has a typical $LOS$ velocity for its environment). The origin of 
VFTS~72 is unclear as its direction of motion is $\sim2\sigma$ 
away from R136.

While the uncertainties on the inferred evolutionary ages of these two stars
are significant, they are systematically lower than their dynamical ages
assuming both stars have been ejected from R136, or close to that cluster. 
As such, both stars, and VFTS~16 in particular, are useful constraints on stellar
evolution models of very  massive stars.

Finally, assuming VFTS~16 was ejected from R136, the
dynamics of this star set a robust lower limit on the age of the cluster 
of $\sim$1.3 Myr.

\begin{acknowledgements}
This work has
made use of data from the ESA space mission Gaia
(http://www.cosmos.esa.int/gaia), processed by the
Gaia Data Processing and Analysis Consortium (DPAC,
http://www.cosmos.esa.int/web/gaia/dpac/consortium).
Funding for the DPAC has been provided by national
institutions, in particular the institutions participating
in the Gaia Multilateral Agreement. DJL thanks
Alex Bombrun and Jose Hernandez of the ESAC {\em Gaia} Science
Operations Centre, and Nate Bastian, for many useful discussions. 
SdM has received funding under the European Unions Horizon 2020 research and innovation programme from the European Research.
AH and S.S.-D. acknowledge financial support from the Spanish Ministry of Economy and Competitiveness (MINECO) through grants AYA2015-68012-C2-1 and Severo Ochoa SEV-2015-0548, and grant ProID2017010115 from the Gobierno de Canarias. 
This research made use
of Simbad and Vizier provided by CDS, Strasbourg;  ESASky, developed by the ESAC
 Science Data Centre (ESDC); and
 TOPCAT. We also thank the HSTPROMO (High-resolution Space Telescope PROper MOtion) Collaboration at STScI for the sharing of their ideas and software.
 
 Finally, our colleague Nolan Walborn, who passed away earlier
 this year, was a strong advocate of the idea that the two
 stars discussed here were likely ejected from R136. He would have welcomed 
 Figure 4 with his usual enthusiasm!

\end{acknowledgements}

%
%

\end{document}